%Paper: hep-th/9207033
%From: physth@ulb.ac.be
%Date: Fri, 10 Jul 92 16:52:54 +0200

%234567890123456789012345678901234567890123456789012345678901234567890123456
%one postscript figure after tex file
%%%%%%%%%%%%%%%%%%%%%%%%%%%%%%%%%%%%%
\def\sqr#1#2{{\vcenter{\vbox{\hrule height.#2pt
          \hbox{\vrule width.#2pt height#1pt \kern#1pt
           \vrule width.#2pt}
           \hrule height.#2pt}}}}
\def\square{\mathchoice\sqr68\sqr68\sqr{4.2}6\sqr{3}6}
\def\lrpartial{\mathrel{\partial\kern-.75em\raise1.75ex\hbox
{$\leftrightarrow$}}}
\def\p{\prime}

\RREF\unruh{W.G. Unruh, Phys. Rev. D {\bf 14}, 870 (1976).}

\RREF\unruhwald{W.G. Unruh, R.M. Wald, Phys. Rev. D {\bf 29}, 1047
(1984).}

\RREF\raine{D.J. Raine, D.W. Sciama, P.G. Grove, Proc. R. Soc. Lond. A
(1991) 435, 205-215.}

\RREF\unruhzureck{W.G. Unruh, W.H. Zurek, Phys. Rev. D {\bf 40} 1071
(1989).}

\RREF\hinter{F. Hinterleitner, University of Vienna preprint, UWThPh -
1991 - 22.}

\RREF\grove{P.G. Grove, Class. Quant. Grav. 3, 801 (1986).}

\RREF\previous{For previous work on this or related models see W.G.
Unruh and W.H. Zurek\attach{\lbrack 7 \rbrack} and references theirin;
see also references cited in F Hinterleitner\refmark{\hinter}. }
\pubnum{ULB-TH-03/92}
\date{July 1992}
\titlepage
\title{ On the Problem of the Uniformly Accelerated Oscillator}
\author{S. Massar, R. Parentani, R. Brout \footnote{*}{e-mail:
ulbg062@bbrnsf11.bitnet}} \address{Service de Physique Theorique,
Universite Libre de Bruxelles \break Campus Plaine, C.P. 225, Bd du
Triomphe, B-1050 Brussels, Belgium } \abstract The particle detector
model consisting of a harmonic oscillator  coupled to a scalar field in
$1+1$ dimensions is investigated in the inertial case.  The same
approach is then used in the accelerating case.  The absence of
radiation from a uniformly accelerated detector in a stationnary state
is discussed and clarified.  \endpage

\noindent 1. Introduction

An interesting exchange of ideas has developed in the  literature
concerning the problem of the accelerated observer in Minkowski
vacuum.  As is well known, dynamical processes get set up which ensure
that the  accelerator be maintained in a thermal distribution
characterized by temperature  $T = (a/2\pi)$, $a$ being the
acceleration taken (necessarily) to be constant\refmark{\unruh}.  These
processes are accompanied by transitions of the radiation field to
which the accelerator is coupled.  The question is : what, in fact,
does the radiation field do during these radiative
events\refmark{\unruhwald}\refmark{\grove} ?  The naive (and incorrect)
answer is that photons are always emitted whether or not the
accelerator is excited or de-excited, since, after all, the vacuum can
only be excited. The correct answer was given by  Grove\refmark{\grove}
who argued that in the presence of constant acceleration a stationary
state is set up in the radiation field as well as in the accelerator in
which there is never emission to infinity, but only fluctuations.  We
shall call this Grove's theorem.  This is in complete analogy to an
atom in a thermal black body.  The implication is then, that a
polarization cloud surrounds the accelerator at all times and energy
gets exchanged with it - locally.  There is no restoration of energy
balance by appealing to photon emission at infinity.

A detailed account of how Grove's theorem works is the recent
contribution of Raine, Sciamia and Grove (RSG)\refmark{\raine}.  The
model considered is that of a point-like oscillator linearly coupled to
a radiation field.  It is exactly solvable, but the answers one gets are
sufficiently abstruse in character that one - or at least we - feels
the need to make contact with more familiar physical concepts.  Such is
the purpose of the present paper.

In Section 2, we first analyze the case of the inertial oscillator.  It
is completely trivial to write down the Green's function of the
oscillator coordinate, $q$, and one comes up with the surprising result
that the self-energy is pure imaginary.  One might leave it at that,
but for the fact that it is precisely this fact that plays a crucial
role in RSG's realization of Grove's theorem.  So physical
interpretation becomes most desirable.  We show how this imaginary self
energy is related to the decay width of the excited states of the
oscillator.  To order $e^2$  it is the width of the first excited
state.  At the same time we show how it also  encodes the level shift of
the ground state.  Continuing the analysis of the  inertial case, we
proceed to show how a local polarization cloud gets built up around the
oscillator.

Section 3 begins with the announcement of Grove's theorem and  our
paraphrasing of his argumentation is tantamount to a proof.  Finally we
give a greatly shortened version of RSG which we believe is somewhat
more transparent than theirs.

\noindent 2. The inertial case

The model of RSG is described by the action in $1+1$
dimensions\refmark{\previous}
$$S = \int dxdt \left\{ {1 \over 2}
[(\partial _t \phi)^2 - (\partial_x \phi)^2] + \delta (x) \left[ {m
\over 2} \left({dq \over dt} \right)^2 - {m \omega_0^2 \over 2} q^2 +
e{dq \over dt} \phi \right] \right\}
\eqn\eqi$$
The propagator of $q$ is found
to be
$$\int dt G_F (t) e^{i \omega t} = {1 \over m \left(\omega^2 -
\omega_0^2 + i {e^2 \over 2 m}  | \omega |\right)}
\eqn\eqii$$
where $G_F (t) = \bra{0} {Tq(t) q(0)}\ket{0}$ and $\ket{0}$
means ground state
oscillator and vacuum of $\phi$ quanta.
We indicate two derivations of \eqii\  :
first through the Heisenberg equations of
motion because of its usefulness in the
accelerating case and the second through
functional integration because of its
immediate simplicity.

The Heisenberg equations of motion are
$$\eqalignno{& \square \phi = e \dot q \delta (x) & \eqname\eqiii \cr
& m [\ddot q + \omega_0^2 q] = -e \dot \phi & \eqname\eqiv \cr}$$
Eq. \eqiii\  solves to
$$\eqalign{\phi &= \phi^0 + \int dt^\p dx^\p G_{ret}
(t,x;t^\p x^\p) e \dot q \delta (x^\p) \cr
&= \phi^0 + {e \over 2} \theta (x) q (t-x) +
{e \over 2} \theta (-x) q (t+x)
\cr}\eqn\eqv$$
where $\phi^0$ is the free-field operator
and we have used the explicit 1
dimensional retarded propagator of a massless field
$$G_{ret} = {1 \over 2} \theta (-t-x)  \theta (-t +x) \eqn\eqvi$$

Substituting \eqv\  into \eqiv\  gives
$$m \ddot q + m \omega_0^2 q + {e^2 \over 2} \dot q = - e
\dot \phi^0 \eqn\eqvii$$
which solves to
$$q (t) = q_0 (t) - \int_{-\infty}^{+ \infty}
dt^\p \chi (t-t^\p) e \dot \phi^0
(t^\p) \eqn\eqviii$$
where
$$\eqalign{& \chi (t) = \theta (-t) {-i \over m
(\omega_1 - \omega_2)} (e^{-i
\omega_1 t} - e^{-i \omega_2 t} ) \cr
& q_0 (t) = A e^{-i \omega_1 t} + B e^{-i \omega_2 t} \cr}
\eqn\eqix$$
and $\omega_1, \omega_2$ = the two roots of
$m (\omega^2 - \omega_0^2) + i
e^2 \omega / 2 = 0$. (They have negative imaginary part.)

The physical solutions are thus damped and
asymptotically the homogeneous term
$q_0 (t)$ in \eqviii\  tends to zero.
{}From now on it is dropped.  Taking Fourier
transforms
$$\eqalignno{& q_\omega = i e \omega \chi_\omega (\phi_R^0 (\omega) +
\phi_L^0 (\omega)) & \eqname\eqx \cr
& \chi_\omega = ( -m \omega^2 + m \omega_0^2 - i {e^2 \omega /2} )^{-1}
& \eqname\eqxi
\cr}$$ and
$$\eqalign{\phi_R^0 (\omega) &=
{1 \over \sqrt{4 \pi \omega}} a_R (\omega) ;
\omega > 0 \cr
& = {1 \over \sqrt{- 4 \pi \omega}} a_R^+ (- \omega) ; \omega < 0 \cr
\phi_L^0 (\omega) &= {1 \over \sqrt{4 \pi \omega}} a_L (\omega) ;
\omega > 0 \cr
&= {1 \over \sqrt{- 4 \pi \omega}} a_L^+ (- \omega) ; \omega < 0 \cr}
\eqn\eqxii$$
where $R,L$ correspond to right, left movers respectively, i.e.
$\phi^0 (t,x) =
\phi_R^0 (u) + \phi_L^0 (v)$ where $u,v = t \mp x$.
Upon substituting \eqxii\
into \eqx, squaring and taking vacuum expectation values, one
obtains
$$\bra{0} {q (t) q (0)} \ket{0} = \int_0^{+\infty} {d \omega \over 2
\pi} e^{-i \omega t} e^2 \omega | \chi_\omega |^2
\eqn\eqxiii$$
Taking the Fourier transform of
$\theta (t) \bra{0} {q(t) q(0)} \ket{0} +
\theta (-t) \bra{0}{q(0)q(t)} \ket{0}$ yields \eqii\  as announced.

The same result follows from functional integration.  One integrates
$e^{i S}$
over the configurations of $\phi$ to find the effective action of $q$
$$S_{eff} (q) = \int {dt \over 2}
[m (\dot q^2 - \omega_0^2 q^2) + e^2 \int
{dtdt^\p \over 2} \dot q (t) D_F^0 (t-t^\p) \dot q (t^\p)
\eqn\eqxiv$$
where $D_F^0$ is the unperturbed Feynman propagator of the
$\phi$ field at
$\Delta x = 0$.  Its Fourier transform is
immediately calculated to be
$-i/2 | \omega |$ whereupon \eqii\
follows by functional integration over
$q$.

It is not without interest to note that the agreement between these two
calculations stems from the asymptotic neglect of the transient $q_0
(t)$ of Eq. \eqviii\ , an effect which is due to the  instability of
excited states (i.e. the damping solutions encoded in $Im \omega_{1,2}$
of  Eq. \eqix .  The ground state being stable is of course
characterized by the functional integration $i \epsilon$ trick which is
used to get \eqxiv .   In short, the damping effect in the solutions of
the equations of motion leads  to ground state expectation values in a
manner equivalent to usual functional integration.

It has been our experience in discussions with colleagues that there
reigns a certain state of confusion regarding the propagator \eqii .
Atomic physicists are accustomed to working with propagation of states
(e.g. $\bra{0}{e^{-i Ht}} \ket{0}$).  Because in the present problem
the ground state is stable the F.T. of such a propagator will have a
pole at real rather than complex frequency, say at $\omega = \omega_0/2
+ \Delta$ where $\Delta$ is the level shift of the ground state.  So
the first question we address is the relation of the Feynman propagator
of $q (t)$ to more familiar atomic concepts.

Firstly note that the perturbative sequence that gives rise to Eq
\eqii\ is given by the integral over intermediate times  $t_1 \ldots
t_i \ldots$ of Fig. 1 wherein the self energy in the $i$th step is $e^2
D_F^0 (t_{i+1} - t_i)$.  Actually if one respects time orderings Fig. 1
is a sum over all possible orderings, hence containing ``Z-graphs'' as
indicated in Fig.2.  The correspondence with (virtual) states  in
perturbation theory is that a graph with $n$ solid lines corresponds to
the $n$th oscillator level.   Thus in Fig.2, for $t<t_0$ and $t>t_f$,
the oscillator is in its ground state.   Contact is made with the
lowest order of usual perturbation theory by stopping  at second order.
The graph with no $Z$ contributions (Fig. 3) includes states $n=0$,
$n=1$ only.  The self energy from this graph is thus to be identified
with the perturbed energy of the first excited state.  It is complex
and the imaginary part gives the width of the $n=1$ state.  Successive
iterations exponentiate to give
$$|\bra{0} { T q(t) q(0)}\ket{0}| = {1 \over 2 m \omega_0 } e^{-
(\Gamma/2) |t| } \eqn\eqxv$$
$\Gamma/2$ is the width of the first excited state given in perturbation
theory by the golden rule formula
$$\Gamma = e^2 \int_0^\infty d \omega {1
\over 2 \pi \omega}  {\omega_0 \over 2 m}
2 \pi \delta (\omega - \omega_0)
= {e^2 \over 2m} \eqn\eqxvi$$
This value of $\Gamma/2$ agrees with $Im \omega_{1,2}$
for small $e^2$, i.e. \eqxv\
agrees with $\bra{0} {q(t) q(0)} \ket{0}$ evaluated
using the Heisenberg
equations of motion.

It is a more subtle business to get the ground state
level shift.  To this end
note that
$${\partial E \over \partial \omega_0^2} =
{\partial \bra{0}{H}\ket{0} \over
\partial \omega_0^2} = {m \over 2} \bra{0} {q^2} \ket{0}
\eqn\eqxvii$$
hence calculable as the concidence limit of the propagator Eq \eqxiii .
One can then integrate over $\omega_0^2$ to get the energy.

An alternative manner of obtaining \eqxvii\  is
through the path integral, by
taking the derivative of $W$ with respect to $\omega_0^2$ [where
$W(T) = i \ln ( \int {\cal D} q {\cal D}
\phi e^{i S (-T/2 , + T/2)} )$]
$$\eqalignno{{\partial E \over \partial
\omega_0^2} &= \lim_{T \to \infty} {1
\over T} {\partial W (T) \over \partial
\omega_0^2} = {m \over 2} \bra{0} {q^2}
\ket{0} & \eqname\eqxviii \cr
&= {\partial \over \partial \omega_0^2}  \left[{-i \over 2}
\int_{-\infty}^{+\infty} {d \omega \over 2 \pi}
\ln (m \omega^2 - m^2 \omega_0^2
+ i (e^2)/2 |\omega|) \right] \cr
&= {i \over 2} \int_{-\infty}^{+\infty} {d
\omega \over 2 \pi} {1 \over \omega^2 - \omega_0^2 +
i {e^2 \over 2m} |\omega|}
& \eqname\eqixx \cr}$$
In spite of appearences \eqixx\  is pure real
thereby verifying the stability
of the ground state.  Indeed
$$Im \left( \lim_{T \to \infty} {1 \over T} {\partial W (T)
\over \partial
\omega_0^2} \right) = {1 \over 2} \int_{-\infty}^{+\infty} d \omega
{\omega^2 -
\omega_0^2 \over (\omega^2 - \omega_0^2)^2 + \left({e^2 \over 2m}
\right)^2
\omega^2} = 0
\eqn\eqxx$$
(verifiable by summing the residus of the poles after closing
the contour either above or below).
Evaluation of the real part of \eqixx\  (which coincides with \eqxiii\
to irrelevant
factors) is trickier and we carry out the calculation for small
$e^2$.  At
lowest order in $e^2$, one may replace the term in $e^2$
in the denominator of
\eqixx\  by $[e^2/(2m)]^2 \omega_0^2$.  In this approximation
$$Re \left( \lim_{T \to \infty} {1 \over T}
{\partial W (T) \over \partial
\omega_0^2} \right) =
{1 \over 2 \pi i} {1 \over 4
\omega_0} \int_0^\infty d (\omega^2) {1 \over \omega^2 - \omega_0^2 - i
{e^2 \over 2m} \omega_0} - {1 \over \omega^2 - \omega_0^2 + i {e^2
\over 2m} \omega_0} + O(e^4) \eqn\eqxxi$$
Performing the logarithmic integrals and keeping careful
track of the phases
that come from the position of the poles in \eqxxi\  gives
$$\eqalign{{\partial E \over \partial \omega_0^2}
&= {1 \over 2 \pi i} {1 \over
 4\omega_0} \ln \left( {e^{i (\pi - (e^2/2m) (1/\omega_0))} \over
e^{-i (\pi - (e^2/2m) (1/\omega_0))}} \right) + O (e^4) \cr
&= {1 \over 4 \omega_0} - {e^2 \over 2m} {1 \over 4 \pi}
{1 \over \omega_0^2} +
O (e^4) \cr}\eqn\eqxxii$$
(The zeroth order term in \eqxxii\  can be obtained by noting that
$\lim_{e^2 \to 0} {1 \over \omega^2 - \omega_0^2 - i
{e^2 \over 2m} \omega_0}
= {P \over \omega^2 - \omega_0^2 } + i \pi
\delta (\omega^2 - \omega_0^2)$ and
using this in \eqxxi.)

Integrating over $\omega_0^2$ between finite limits $\omega_1^2$ and
$\omega_2^2$ gives
$$E^{(2)} - E^{(1)} = \left({1 \over 2} \omega_1 -
{1 \over 2 \pi} {e^2 \over
2m} \ln \omega_1 \right) - \left({1 \over 2} \omega_2 - {1 \over 2 \pi}
{e^2 \over 2m} \ln \omega_2 \right) + O (e^4)\eqn\eqxxiii$$
This result is the same as that obtained by usual perturbative theory
$$\eqalign{E &= {1 \over 2} \omega_0 + {e^2 \over 2m}
\int_0^\infty {d \omega
\over 2 \pi \omega} - {e^2 \over m^2} \int_0^\infty
{d \omega \over 2 \pi
\omega} {1 \over \omega_0 + \omega} {m \omega_0 \over 2} \cr
&= {1 \over 2} \omega_0 + {1 \over 2 \pi} {e^2 \over 2m} \int_0^\infty
{d \omega
\over \omega + \omega_0} = {1 \over 2} \omega_0 -
{1 \over 2 \pi} {e^2 \over 2m}
\ln {\omega_0 \over \Lambda} \cr}\eqn\eqxxiv$$
where $\Lambda$ is an ultraviolet cut-off.  In Eq. \eqxxiv\  there are
two terms in the perturbation, first order in the ``seagull'' $(e^2
\phi^2)$ and second order in $e \dot q \phi$ (note that the seagull
term is necessary for compatibility with \eqi\ ) obtained by going into
the  canonical formalism, here through use of the momentum $p = m (\dot
q + e \phi)$.

The above gives one some feeling for the information encoded in $G_F$.
We now explore the configuration of the field $\phi$.   From \eqv\  we
have
$$\eqalign{\langle \phi (t,x) \phi (t^\p, x^\p) \rangle
&= \langle \phi^0 (t,x)
\phi^0 (t^\p, x^\p) \rangle + {e^2 \over 4}
\langle q (t+x) q (t^\p + x^\p)
\rangle \cr
& + {e \over 2} \langle \phi^0 (t,x) q(t^\p + x^\p)
\rangle + {e \over 2}
\langle q (t+x) \phi^0 (t^\p, x^\p) \rangle \cr}
\eqn\eqxxv$$
where for definiteness we have taken both $x, x^\p < 0$.
Using equation  \eqx\  we
get
$$\eqalign{\langle \phi (t & ,x) \phi (t^\p, x^\p) \rangle -
\langle \phi^0 (t,x)
\phi^0 (t^\p, x^\p) \rangle \cr
& = {e^4 \over 4} \langle \int d \omega e^{-i \omega v}
\omega \chi_\omega
(\phi_R^0 (\omega) + \phi_L^0 (\omega))
 \int d \omega^\p e^{i \omega^\p v^\p} \omega^\p \chi_{\omega^\p}^*
(\phi_R^{0 +} (\omega^\p) + \phi_L^{0+} (\omega^\p)) \rangle \cr
 &\  +i{e^2 \over 2} \langle \int d \omega e^{-i \omega v}
\omega \chi_\omega
(\phi_R^0 (\omega) + \phi_L^0 (\omega)) (\phi_R^0 (u^\p) +
\phi_L^0 (v^\p))
\rangle \cr
 &\  - i {e^2 \over 2} \langle (\phi_R^0 (u) + \phi_L^0 (v))
\int d \omega^\p
e^{i \omega^\p v^\p} \omega^\p \chi_{\omega^\p}^*
(\phi_R^{0+} (\omega^\p) +
\phi_L^{0+} (\omega^\p)) \rangle \cr}\eqn\eqxxvi$$
where we have used the fact that $\phi_R^0$ ($\phi_L^0$) depend on $u$
($v$) only . When the argument of these fields is $\omega$, it means
the Fourier transform with respect to $u$ or $v$ respectively, as in Eq
\eqxii .

Regrouping terms in judicious fashion and writing everything in terms
of the configuration space variables $u$ and $v$ gives for the r.h.s.
of Eq \eqxxvi\ ) $$\eqalign{& \biggl\lbrack \int \int { d \omega d
\tilde v \over 2 \pi } e^{-i \omega ( v-\tilde v )}
\int \int { d \omega^\p d \tilde v^\p \over 2 \pi }
e^{-i \omega^\p ( v^\p
-\tilde v^\p )}\cr
&\quad\quad \Bigl \lbrack {e^2 \over 2} \omega
\omega^\p \chi_\omega
\chi_{\omega^\p}^* (\langle \phi_R^0 (\tilde v)
\phi_R^{0} ({\tilde  v}^\p)
\rangle + \langle \phi_L^0 (\tilde v) \phi_L^{0}
({\tilde  v}^\p) \rangle)
+ (i  \omega \chi_\omega - i  \omega^\p
\chi_{\omega^\p}^*) \langle \phi_L^0 (\tilde v) \phi_L^{0} ({\tilde
v}^\p) \rangle
\Bigr \rbrack \biggr\rbrack \cr
&\ + i{e^2 \over 2} \int \int { d \omega d \tilde v \over 2 \pi }
e^{-i \omega ( v-\tilde v )} \omega \chi_\omega
\langle \phi_R^0 (\tilde v)
\phi_R^0 (u^\p) \rangle \cr
&\ - i{e^2 \over 2} \int \int { d \omega^\p d
{\tilde v}^\p \over 2 \pi }
e^{-i \omega^\p ( v^\p-{\tilde v}^\p )} \omega^\p
\chi_{\omega^\p}^*
\langle \phi_R^0 (u) \phi_R^0 ( {\tilde v}^\p) \rangle \cr}
\eqn\eqxxvii$$

The integrand of \eqxxvii\  has been written as a sum of six terms in
which the first four have been grouped together. These first four terms
in fact cancel.  This follows from the use of translational invariance
of all expectation values so that the integrand is a function of
$(\tilde v -{\tilde v}^\p)$ only, whereupon integration over $(\tilde v
+{\tilde v}^\p) / 2$ gives $\delta ( \omega - \tilde \omega )$. Use of
$$Im \chi_\omega = {e^2 \over 2} \omega |\chi_\omega |^2
\eqn\eqxxviii$$
as follows from Eq \eqxi\  then gives
the abovementioned cancellation. There
results the extremely simple expression
$$\langle \phi (t,x) \phi (t^\p, x^\p)
\rangle - \langle \phi^0 (t,x)
\phi^0 (t^\p, x^\p) \rangle =  i
{e^2 \over 8 \pi} \int_0^\infty d \omega \{
e^{-i \omega (v-u^\p)}  \chi_\omega -
e^{i \omega (u-v^\p)}  \chi_\omega^* \}
\eqn\eqixxx$$

Before exploring the physics implied by \eqixxx\   we comment on the
fundamental relation \eqxxviii\  which is of the form of a
fluctuation-dissipation theorem in statistical mechanics, and related
to the unitarity condition in quantum mechanics such as encountered in
the Breit-Wigner formula. In fact it follows here from the canonical
structure of the theory. From the form  $\phi = \phi^0 - e^2 G_{ret}
\dot \chi \phi^0$ (Eq \eqv\  and \eqx\ ) one computes  $\lbrack \phi
(t,x) , \phi (t, x^\p) \rbrack$. Expressing the result in Fourier
components, one checks that the necessary condition for the vanishing
of the equal time  commutator for all $x$, $ x^\p$ is Eq \eqxxviii\ .
The use of  the canonical structure of the theory clearly closely
parallels the more usual appeal to unitarity to yield the optical
theorem wherein the square of the scattered wave is related to the
amount of amplitude taken out of the forward wave.

The first thing to remark is that Eq. \eqixxx\  implies $\langle T_{\mu
\nu} \rangle = 0$ as is seen directly from
$$\langle T_{uu} \rangle - \langle T_{uu} \rangle_0 =
\lim_{\scriptstyle u^\p
\to u \atop \scriptstyle v^\p \to v} \partial_u \partial_{u^\p}
 \lbrack \langle \phi (u,v)
\phi (u^\p, v^\p) \rangle - \langle \phi^0 (u,v)
\phi^0 (u^\p,v^\p) \rangle
\rbrack\eqn\eqxxx$$
Similarily $\langle T_{vv} \rangle - \langle T_{vv} \rangle_0 = 0$
. Finally
$\langle T_{uv} \rangle = 0 $ since we are dealing with a
massless field in
$2$ dimensions and  $T_{uv} = {1 \over 4} tr T$. Since the
curvature is zero
there is no trace anomaly.

Nonetheless there is a polarization cloud set up around
the oscillator as is
seen from
$$\eqalign{\langle  \left( \phi (t,x) \right) ^2 \rangle -
\langle \left( \phi^0 (t,x) \right) ^2 \rangle
& = {e^2 \over 8 \pi}
\int_0^\infty d \omega \left ( i \chi_\omega e^{+2i \omega |x|}
\chi_\omega +
c.c.\right ) \cr
& \simeq  {1 \over 2 \omega_0} \left({e^2 \over 2m} \right)
 e^{-(e^2/2m)
|x|} \cos (2 \omega_0 x) \cr}\eqn\eqxxxi$$
Thus it is static and carries no energy in the mean. It is this
dressing which signals the fact that the true state is a superposition
of free oscillator and radiation states, virtual photons emitted and
absorbed by the oscillator. These get distorted by the acceleration but
nevertheless continue to remain local, as we shall now see.

\noindent 3. The Accelerated Oscillator

The non-radiation of the accelerating observer (with constant
acceleration) is an essential remark due to Grove.  Reference \grove\
contains the germ of his argumentation.  In the foregoing paragraphs we
attempt to formalize these considerations so as to justify the
appelation ``Grove's Theorem'' in referring to his contribution.

The upshot of the matter is the exploitation of the invariance of the
Minkowski vacuum state under boosts in conjunction with the
conservation of energy momentum, $T_{\mu \nu}$, to wit:

Given:
\item{a)} A Poincare invariant field.
\item{b)} A detector following a uniformaly accelerated trajectory in
flat space $\rho = a^{-1}$ (where $\rho, \tau$
are Rindler coordinates defined
by $x = \rho \cosh a \tau, t = \rho \sinh a \tau$).
\item{c)} A $\tau$ independent coupling.

Then:
\item{1)} The operator $T_{\mu \nu}$
(the stress tensor of the total system:
field plus detector plus interaction)
does
not depend explicitly on $\tau$.  This is because $\partial_\tau$ is a
Killing vector of flat space i.e. the metric is invariant under
translations of $\tau$ (boosts).
\item{2)} Postulating that the interaction gives rise to a sort of
ergodicity, at least in the weak form, the system forgets initial
conditions. Since the vacuum is invariant under boosts, the system must
arrive at a stationary state independent of $\tau$.
\item{3)}  There is no outgoing flow as is seen from the following
construction. Draw a box surrounding a portion of the trajectory of the
accelerator, bounded by the lines $\rho = a_1^{-1}$, $\rho = a_2^{-1}$
where $a_1^{-1} < a^{-1} < a_2^{-1}$ and $\tau = \tau_1, \tau =
\tau_2$.  Conservation of $T_{\mu \nu}$ tells us that
$$ \int_{\tau_1}^{\tau_2} d \tau
\langle T^\tau_\rho|_{\rho = a_1^{-1}} -
T^{\tau}_\rho|_{\rho = a_2^{-1}}
\rangle = \int_{a_1^{-1}}^{a_2^{-1}}
\rho d \rho \langle T_\tau^\tau|_{\tau =
\tau_2} - T_{\tau}^\tau|_{\tau = \tau_1} \rangle
\eqn\eqxxxii$$
{}From 1) and 2), the r.h.s. of \eqxxxii\  vanishes. Then since we are
working in Minkowski vacuum the flows $\langle T^\tau_\rho \rangle$
must be directed out from the detector (since in-flow would require the
presence of asymptotic photons flowing in from infinity and these are
absent in vacuum).  Excluding the possibility of eternal negative
energy flow (only local negative energy flow can be constructed from
coherent field states), the two terms on the l.h.s. of \eqxxxii\ are
both positive and add to zero. They are thus separately zero : QCD.
 An alternative line of reasoning is by appeal to parity invariance.
Suppose the detector does not distinguish between right and left movers
in its interaction. Then because Minkoxski vacuum is symmetric under
the interchange between the left and right pieces of the radiation
field in Rindler coordinates the two out-flows $\langle
T_\tau^\rho|_{\rho = a_1^{-1}}\rangle$ and $\langle
T_{\tau}^\rho|_{\rho = a_2^{-1}}\rangle$ of Eq \eqxxxii\  are equal.
Since their sum is zero they are seperately zero.

We now show how in the RSG model, Grove's theorem is realized, by
essentially repeating the analysis of RSG in a highly synthetic fashion
so as to bring out the fact that the non-radiation is essentially an
expression of the same mathematics as in the inertial case (Section
2).  This is in line with the fact that the above proof of Grove's
theorem is a straightforward generalization of the fact that an
inertial system does not radiate forever.

The action for the uniformly accelerated detector taken in the right
Rindler quadrant is
$$S = \int dx dt {1 \over 2} [(\partial_t \phi)^2 - (\partial_x
\phi)^2] + \int d \rho d \tau \delta
(\rho - a^{-1})  \left[ {m \over 2}
\left({dq \over d \tau} \right)^2 +
{m \over 2} \omega_0^2 q^2 + e {dq \over d
\tau} \phi (\tau, a^{-1}) \right] \eqn\eqxxxiii$$
to give rise to the Heisenberg
equations of motion
$$\eqalign{& \square \phi =  e {dq \over d \tau} \delta
(\rho - a^{-1}) \cr & m \left({dq \over d \tau}
\right)^2 + m \omega_0^2 q^2 =
-e {d \phi (\tau, a^{-1}) \over d \tau} \cr}
\eqn\eqxxxiv$$
which may be integrated as in the inertial case to
$$\phi (x,t) = \phi^0 (x,t) + {e \over 2} q
(\tau_{ret})\eqn\eqxxxv$$
($\tau_{ret}$ is the value of $\tau$ at the
intersection of the past light cone
from $(x,t)$ with the trajectory)
$$\eqalignno{& q (\tau) = q_0 (\tau) + i e
\int d \lambda e^{-i \lambda \tau}
\lambda \chi_\lambda [\phi^0_R (\lambda) +
\phi^0_L (\lambda)] & \eqname\eqxxxvi \cr
& \chi_\lambda = ( -m \lambda^2 + m \omega_0^2 - i e^2 / 2
\lambda)^{-1} & \eqname\eqxxxvii \cr }$$
and
$$\eqalign{ & \phi^0_R (\lambda) = \int {d \tau \over 2
\pi} e^{i \lambda \tau} \phi^0_R (-
e^{-a \tau}/a) \cr
& \phi^0_L (\lambda) = \int {d \tau \over 2
\pi} e^{i \lambda \tau} \phi^0_L (
e^{a \tau}/a) \cr}\eqn\eqxxxviii$$
where we have used that $u=-a^{-1} e^{-a \tau}$ and $v=a^{-1}
e^{a \tau}$ in Rindler coordinates.
The transient $q_0$ is once more asymptotically damped to zero. The
oscillator Green's function is
$$\langle q (\tau) q (0) \rangle =  e^2 \int_0^\infty
{d \lambda \over  \pi} e^{-i \lambda \tau}  \lambda |\chi_\lambda|^2
({1 \over 2} + {1 \over e^{2 \pi \lambda / a} - 1 })\eqn\eqixxxx$$
where we recall that the Green's function of
$\phi_0$ in Rindler coordinates is thermal
with temperature $T = (a/2\pi)$ :
$$\langle a_\lambda^+ a_\lambda \rangle =
(e^{(2 \pi / a) \lambda} -1)^{-1}
\eqn\eqxxxx$$
($a_\lambda^+$ creates a Rindleron, $a_\lambda$ destroys one).

Similarly the field propagator follows
the same pattern as in \eqxxv\  to
give
$$\eqalign{\langle \phi (t,x) &\phi
(t^\p, x^\p) \rangle = \langle \phi^0
(t,x) \phi^0 (t^\p, x^\p) \rangle +
{e^2 \over 4} \langle q (\tau_{ret}) q
(\tau_{ret}^\p) \rangle \cr
& + {e \over 2} \langle \phi^0 (t,x)
q (\tau^\p_{ret}) \rangle + {e \over 2}
\langle q (\tau_{ret}) \phi^0 (t^\p , x^\p) \rangle \cr}
\eqn\eqxxxxi$$
where for definitesness we have taken
$(t,x), (t^\p, x^\p)$ to the left of the
trajectory and $\tau_{ret} = a^{-1} \ln (va) $,
$ \tau_{ret}^\p = a^{-1} \ln
(v^\p a)$.  The fluctuation due to the interaction,
using the same procedure as
that of Eqs \eqxxvi\ \eqxxvii , is

$$\eqalignno { \langle \phi (t & ,x) \phi (t^\p, x^\p)
\rangle - \langle \phi^0
(t,x) \phi^0 (t^\p, x^\p) \rangle \cr
 & = {e^4 \over 4} \langle \int d \lambda
e^{-i \lambda \tau_{ret}}  \lambda \chi_\lambda
(\phi_R^0 (\lambda) + \phi_L^0 (\lambda))
 \int d \lambda^\p e^{i \lambda^\p \tau_{ret}^\p}
\lambda^\p \chi_{\lambda^\p}^*
(\phi_R^{0 +} (\lambda^\p) + \phi_L^{0+} (\lambda^\p)) \rangle \cr
 & \ +i{e^2 \over 2} \langle \int d \lambda
e^{-i \lambda \tau_{ret}} \lambda \chi_\lambda
(\phi_R^0 (\lambda) + \phi_L^0 (\lambda))
(\phi_R^0 (u^\p) + \phi_L^0 (v^\p))
\rangle \cr
 &\  - i {e^2 \over 2} \langle (\phi_R^0 (u) +
\phi_L^0 (v)) \int d \lambda^\p
e^{i \lambda^\p \tau_{ret}^\p} \lambda^\p
\chi_{\lambda^\p}^*   (\phi_R^{0+} (\lambda^\p) +
\phi_L^{0+} (\lambda^\p)) \rangle & \eqname\eqxxxxii \cr
& = \biggl\lbrack \int \int { d \lambda d \tilde \tau \over 2 \pi }
e^{-i \lambda ( \tau_{ret} -\tilde \tau )}
\int \int { d \lambda^\p d \tilde \tau^\p \over 2 \pi }e^{-i \lambda^\p
(\tau_{ret}^\p -\tilde\tau^\p )}\cr
& \quad \quad \Bigl \lbrack {e^2 \over 2}
\lambda \lambda^\p \chi_\lambda \chi_{\lambda^\p}^*
(\langle \phi_R^0 (-e^{-a \tilde \tau}/a) \phi_R^{0} (- e^{-a \tilde
\tau^\p}/a) \rangle
+ \langle \phi_L^0 ( e^{a \tilde \tau} /a) \phi_L^{0}
( e^{a \tilde \tau^\p}/a) \rangle) \cr
& \quad\quad + (i  \lambda \chi_\lambda - i  \lambda^\p
\chi_{\lambda^\p}^*) \langle \phi_L^0 ( e^{a \tilde \tau}/a)
\phi_L^{0} ( e^{a \tilde \tau^\p}/a) \rangle
\Bigr \rbrack \biggr\rbrack \cr
&\ + i{e^2 \over 2} \int \int { d \lambda d \tilde \tau \over 2 \pi }
e^{-i \lambda ( \tau_{ret}-\tilde \tau )} \lambda \chi_\lambda \langle
\phi_R^0 (- e^{-a \tilde \tau}/a)
\phi_R^0 (u^\p) \rangle \cr
&\ - i{e^2 \over 2} \int \int { d \lambda^\p d
{\tilde \tau}^\p \over 2 \pi }
e^{-i \lambda^\p (\tau_{ret}^\p-{\tilde \tau}^\p )}
\lambda^\p \chi_{\lambda^\p}^*
\langle \phi_R^0 (u) \phi_R^0 ( - e^{-a \tilde \tau^\p}/a) \rangle
 & \eqname\eqxxxxiii \cr}$$

In the present case the translational symmetry of the inertial Green's
functions is replaced by the fact that $\langle \phi^0
( e^{a \tau}/a, -  e^{-a \tau}/a) \phi^0 (
e^{a \tau^\p}/a, - e^{-a \tau^\p}/a) \rangle$ is a
function of $(\tau -
\tau^\p )$ only. Thus integration over $(\tau +
\tau^\p )$ in Eq \eqxxxxiii\  gives
$\delta ( \lambda - \lambda^\p )$. The unitarity
relation Eq \eqxxviii\  is true in the present case with $\omega$
replaced by $\lambda$ since the latter labels the frequency in the
proper time of the oscillator and only thus enters  into the definition
of the oscillator's response function. Therefore the same
simplification as in the passage from Eq \eqxvii\  to Eq \eqixxx\
obtains to give (after evaluating the propagators)
$$\eqalign {  \langle \phi (t ,x)
\phi ( t^\p , x^\p ) \rangle - \langle \phi^0
(t,x) & \phi^0 (t^\p , x^\p ) \rangle =
{ i e^2 \over 8 \pi } \int d \lambda\cr
  \biggl \{ (av)^{-i\lambda / a}
|a u^\p|^{-i\lambda / a} \chi_\lambda
& \left \lbrack \theta (u^\p )
{ 2 \over sinh( \pi \lambda / a ) }
+ \theta (-u^\p ) ( {\theta (\lambda )
\over 1 - e^{-2 \pi \lambda / a} }
- {\theta (-\lambda ) \over e^{2 \pi \lambda / a} - 1 } )
\right \rbrack \cr
- (av^\p )^{i\lambda / a}
|a u|^{i\lambda / a} \chi_\lambda^*
&\left \lbrack \theta (u )
{ 2 \over sinh( \pi \lambda / a ) }
+ \theta (-u ) ( {\theta (\lambda )
\over 1 - e^{-2 \pi \lambda / a} }
- {\theta (-\lambda ) \over e^{2 \pi \lambda / a} - 1 } )
\right \rbrack
\biggr \} \cr }\eqn\eqxxxxiv$$

By direct differentiation one checks out Grove's theorem
$\langle T_{uu} \rangle = \langle T_{vv} \rangle =0$,
i.e. there is no energy
flow out from the detector.

The polarization cloud (the coincidence limit of \eqxxxxiv\ ) is once
more non-vanishing. It extends into the future Rindler quadrant and in
the right Rindler quadrant, when expressed in Rindler light like
variables $u_{Rindler}=-a^{-1} ln( -au)$, $v_{Rindler}=a^{-1} ln( av)$,
is  identical to the polarisation cloud surrounding the inertial
oscillator in a thermal bath with temperature $a / 2 \pi$. The presence
of this polarisation cloud was missed in RSG. After this work was
completed, we became aware that identical conclusions have been reached
independently by F. Hinterleitner\refmark{\hinter}.

We are grateful to doctor Peter Grove for introducing us to this
problem and for illuminating discussions concerning it.

\REF\unruhzureckii{W.G. Unruh, W.H. Zurek, Phys. Rev. D {\bf 40} 1071
(1989).} \refout

\vskip 2 cm
\centerline{Figure Captions}

\item{Fig 1} The nth order Feynman graph. Solid (wavy) lines are bare
oscillator (photon) propagators. One gets converted into the other at
intermediate times $t_i$.

\item{Fig 2} The same as in Fig 1) but with different time orderings
implicit in the former now explicitly exhibited.

\item{Fig 3} 2nd order non-Z contribution

\end